\documentclass{article}
\usepackage[margin=1.00in]{geometry}
\usepackage{mathtools} % Includes amsmath
\usepackage{amsfonts}
\usepackage{longtable}
\usepackage{stmaryrd}
\usepackage[mathletters]{ucs}
\usepackage[utf8x]{inputenc}
\usepackage{fancyvrb}
\usepackage{listings}
\usepackage{color}
\begin{document}

%\special{papersize=8.5in,11in}
\setlength{\pdfpageheight}{\paperheight}
\setlength{\pdfpagewidth}{\paperwidth}

%\conferenceinfo{CONF 'yy}{Month d--d, 20yy, City, ST, Country} 
%\copyrightyear{20yy} 
%\copyrightdata{978-1-nnnn-nnnn-n/yy/mm} 
%\doi{nnnnnnn.nnnnnnn}

% Uncomment one of the following two, if you are not going for the 
% traditional copyright transfer agreement.

%\exclusivelicense                % ACM gets exclusive license to publish, 
                                  % you retain copyright

%\permissiontopublish             % ACM gets nonexclusive license to publish
                                  % (paid open-access papers, 
                                  % short abstracts)

%\titlebanner{banner above paper title}        % These are ignored unless
%\preprintfooter{short description of paper}   % 'preprint' option specified.

\begin{titlepage}
\title{MoA Interpretation of the Iterative Conjugate Gradient  Method with  Psi Reduction - A Tutorial to teach the Mathematically literate in Linear and Tensor Algebra: Part I}
\author{Lenore M. Mullin and Paul Sebexen\\
}

\maketitle

%\begin{abstract}
%\end{abstract}
\end{titlepage}
\tableofcontents
%\addcontentsline{toc}{section}{References}
\listoffigures
%\listoftables
\newpage
% \category{CR-number}{subcategory}{third-level}

% general terms are not compulsory anymore, 
% you may leave them out
%\terms
%term1, term2
%keywords
%Arrays

%\section{Introduction}
% See the ``Article customise'' template for come common customisations
\newcommand{\cat}{+\!\!\!+}
\newcommand{\Take}{\mbox{Take}}
\newcommand{\take}{\,\triangle\,}
\newcommand{\Drop}{\mbox{Drop}}
\newcommand{\drop}{\,\nabla\,}
\newcommand{\rshp}{\,\widehat{\Rho}\;}
\newcommand{\Rho}{\,\rho\,}
\newcommand{\Tau}{\,\tau\,}
\newcommand{\Dim}{\,\delta\,}
\newcommand{\transpose}{\bigcirc\!\!\!\!\!\backslash\;}
\newcommand{\kron}{\bigcirc\,\!\!\!\!\!\!\times\;}
\newcommand{\Ravel}{\,\mbox{\tt rav}\,}
\newcommand{\Gradeup}{\,\mbox{\tt gu}\,}
\newcommand{\ip}{_+ \bullet _\times}

\newtheorem{definition}{Definition}
\newtheorem{corollary}{Corollary}
\newtheorem{theorem}{Theorem}
\section{Motivation and Introduction}
\subsubsection{Intentions of Tutorial}
It is often difficult to learn new mathematics semantically {\it and} syntactically, even
when there are similarities in the words and meaning when discussed aloud.  The goal
of this document is to facilitate  learning through explainations and definitions relating
our common mathematical knowledge and highlighting what is new. It is meant to be a working document that will evolve based on feedback from target audiences, those mathematically literate in linear and tensor algebra,  those that want to learn MoA, Psi Calculus~\cite{mul88}, and its uses~\cite{mullin-raynolds-book}, those that want
and need the ability to prove a design, either in hardware or software through the ONF,
 Operational Normal Form, and  those wanting to exploit all resources optimlally, especially when Tensor Algebra, i.e. algorithms foundational to their application,
are needed: Knowledge Representation and Modeling, Scientific Computing, Signal Processing, etc.
\subsubsection{Matrix Multiply and Solvers}
\noindent
We aim to solve for $\vec{x}$, a vector,  in 
\begin{equation}
  A \vec{x} = \vec{b}
\end{equation}
where $A$ is symmetric, real, positive-definite, and all variables are conformable for the operaton, i.e. $\vec x$, $\vec b$, and the number of columns in $A$ have the
same number of components. 

\noindent
The {\em matrix-vector product}, like other inner products from linear algebra, can be expressed in MoA as $\ip$, denoting point-wise scalar extension of multiplication over the inner dimension, followed by a single application of additive reduction.

\noindent
We are motivated by a few anamolies in Linear and Tensor Algebra.
From a CMU CS professor's notes on painless-conjugate-gradient~\cite{painless}:
``... A is an n by n matrix, and x and b are vectors, that is, 
n by 1 matrices.
The inner product of two vectors is written $x^Ty$ and represents the scalar
sum $\sum_{i=1}^{n} x_{i}y_{i}$ ... In general, expressions that reduce to
1 by 1 matrices  are treated as scalars.''

\noindent
Although we can build symbolic software systems that apply the same equalities as 
one might do by hand, that implementation is often slow. If we are to have the 
mathematical prowess of tensor algebra {\bf and} the computational/communication prowess needed
for IoT and the complexity of real time systems with knowledge and reasoning one must realize that
it is hard to prove properties of programs if grammars  have anamolies. Ideally, if everything was
a linear, or multilinear operation, reasoning was possible. 
We often forget that what can be done in our minds, on paper, etc. mathematically, 
must now be transformed to an accurate answer done in the most
computationally efficient  way, given a certain set of resources. What we ignore costs computational
resources to analyze and deal with the anamoly. Linear, or multilinear, operations, have the
advantage that everything is mathematical without exception.
In the above case, at one moment it says a vector is an n by 1, then in the summation
it is represented as a vector. They {\bf ARE} different entities.
In a tensor world, a scalar is a 0-dimensional array, a vector is 1-d, and a 1 by 1, is a 2-d
array. MoA and Psi Calculus give us this formality while providing the full tensor algebra we
are familiar with.

\noindent
We study the CG because it  is the most prominent iterative 
method for solving sparse systems of linear equations.
For  dense systems, factor and solve with
back substitution is often best. 
Re complexity, the time spent factoring a dense matrix
 is roughly equivalent to the time spent solving the system iteratively.

\noindent Finallly, we  study the CG because, if it is fundamental to other solvers. We aim to paramaterizie the
variants,  while we  assure flexible use, performance, and accuracy.

\section{An Iterative Solver: The Conjugate Gradient (CG)}

Rewriting the objective with this new notation, we now aim to solve
\begin{equation}
A \ip \vec{x} \equiv \vec{b}
\end{equation}
where $n$ is a positive integer indicating the size of the problem, and the {\em shapes} are denoted as
\begin{equation}
  \Rho A \equiv < n \; n >
\end{equation}
and
\begin{equation}
  \Rho \vec{x} \equiv \Rho \vec{b} \equiv < n >
\end{equation}

\subsection{Base Case}
\begin{figure}[h]
The {\bf pseudocode} gives us two lines setting the initial condition of the working arrays:
\begin{align}
  \mathbf{r}_0 &:= \mathbf{b} - \mathbf{A} \mathbf{x}_0 \\
  \mathbf{p}_0 &:= \mathbf{r}_k^T0 \label{eq:pseudo0}
\end{align}
\caption{Pseudo Code: Base Case}
\end{figure}

\noindent
The base case populated prior to the first iteration can be expressed by first initializing the guess for $\vec{x}$ as a zero vector of shape $< n >$ (could be set to a better guess if available). Then, we rewrite the pseudocode.
\subsubsection{Pseudocode to MoA: Base Case}
In the pseudocode we see there are {\it k} iterations, implying that a matrix with k rows will hold all recurrences. But, it turns out we only need 2 rows
in each matrix used in the recurrence, i.e. the  previous time step, and the present time step.  We keep the same loop bounds, i.e. $0 \leq
i < n$, but we change the loop body from $i$ and $i+1$ to
$i-i$ and $i+1-i$, i.e. 0 and 1. We then create the following matrices {\it X, P} and {\it R}, each a 2 by n. Notice how Expression (\ref{eq:pseudo0}) looks now. The {\bf Transpose} has
been eliminated because the transpose of a vector in MoA is the same vector.
\begin{figure}[h]
\begin{align}
  <0> \psi X &\equiv <0 \; ... \; 0> \\
  <0> \psi P &\equiv <0> \psi R \equiv \vec{b} - A \ip <0> \psi X
\end{align}
\caption{MoA: Base Case}
\end{figure}
%\begin{equation}
%  <0> \psi RS \equiv (\transpose <0> \psi R) \ip <0> \psi R
%\end{equation}

\subsection{Subsequent Iterations}
\begin{figure}[h]
The pseudocode next gives us
\begin{gather}
  k := 0 \\
  \nonumber \textbf{repeat} \\
  \alpha_k := \frac{\mathbf{r}_k^T \mathbf{r}_k}{\mathbf{p}_k^T \mathbf{A} \mathbf{p}_k} \\
  \mathbf{x}_{k + 1} := \mathbf{x}_k + \alpha_k \mathbf{p}_k \\
  \mathbf{r}_{k + 1} := \mathbf{r}_k - \alpha_k \mathbf{A} \mathbf{p}_k \\
  \text{if $\mathbf{r}_{k+1}$ is sufficiently small, then exit loop} \\
  \beta_k := \frac{\mathbf{r}_{k + 1}^T \mathbf{r}_{k + 1}}{\mathbf{r}_k^T \mathbf{r}_k} \\
  \mathbf{p}_{k + 1} := \mathbf{r}_{k + 1} + \beta_k \mathbf{p}_k \\
  k := k + 1 \\
  \nonumber \textbf{end repeat} \\
  \text{Result is $\mathbf{x}_{k + 1}$.}
\end{gather}
\caption{Pseudo Code: Subsequent Iterations}
\end{figure}
\subsubsection{Pseudocode to MoA: Subsequent Iterations}
The iterative solver can be expressed in MoA for a temporal index, $<i>$, where
\begin{equation}
  0 \leq i < n
\end{equation}
effectively replaces the function of $k$ in the pseudocode and allows the procedure to be written using the array equality operator, ``$\equiv$'', without requiring the assignment operator, ``$:=$''. This can be thought of as unrolling the temporal dimension. This subtlety, is why we are making the change in variable name. 

\subsubsection{The Loop Body}
We rewrite the loop body in MoA syntax by:
\begin{enumerate}
  \item Replacing bold vectors with iteration subscripts, with a matrix and an index of that iteration.  Vectors that are really vectors are denoted by the same named variable, e.g. x,  but as vector symbols: $\mathbf{x} \rightarrow \vec{x}$
  \item Replacing subscript indexing with the $\psi$ binary function: $\mathbf{r}_k \rightarrow <i> \psi {R}$
  \item Using transpose unary prefix operator $\transpose$: $\mathbf{r}_k^T \rightarrow \transpose< i> \psi {R}$
  \item Using inner product binary operator $\ip$, as discussed earlier
\end{enumerate}
\subsubsection{Resulting Algorithm}
\begin{figure}
The resulting algorithm is written as
\begin{align}
  <i> \psi \vec{\alpha} &\equiv
    \frac{(\transpose <i> \psi R) \ip (<i> \psi R)}{(\transpose <i> \psi P) \ip A \,\ip (<i> \psi P)} \\
    <i + 1> \psi X &\equiv (<i> \psi X) + (<i> \psi \vec{\alpha}) \times (<i> \psi P) \\
    <i + 1> \psi R &\equiv (<i> \psi R) - (<i> \psi \vec{\alpha}) \times A \,\ip (<i> \psi P) \\
  <i> \psi \vec{\beta} &\equiv
    \frac{(\transpose <i + 1> \psi R) \ip (<i + 1> \psi R)}{(\transpose <i> \psi R) \ip (<i> \psi R)} \\
    <i + 1> \psi P &\equiv (<i + 1> \psi R) + (<i> \psi \vec{\beta}) \times (<i> \psi P) \\
    &\text{Result is $<i + 1> \psi X$.}
\end{align}
\caption{MoA: Subsequent Iterations}
\end{figure}
The vectors $\vec{\alpha}$ and $\vec{\beta}$ can be substituted in for each usage. The expanded algorithm is now
\subsubsection{Substituting for $\vec{\alpha}$ and ${\vec{\beta}}$}
\begin{align}
%  <i> \psi \vec{\alpha} &\equiv
%    \left( \frac{(\transpose <i> \psi R) \ip (<i> \psi R)}{(\transpose <i> \psi P) \ip A \ip (<i> \psi P)} \right) \\
    <i + 1> \psi X &\equiv (<i> \psi X) + \left( \frac{(\transpose <i> \psi R) \ip (<i> \psi R)}{(\transpose <i> \psi P) \ip A \,\ip (<i> \psi P)} \right) \times (<i> \psi P) \\
    <i + 1> \psi R &\equiv (<i> \psi R) - \left( \frac{(\transpose <i> \psi R) \ip (<i> \psi R)}{(\transpose <i> \psi P) \ip A \,\ip (<i> \psi P)} \right) \times A \,\ip (<i> \psi P) \\
%  <i> \psi \vec{\beta} &\equiv
%    \left( \frac{(\transpose <i + 1> \psi R) \ip (<i + 1> \psi R)}{(\transpose <i> \psi R) \ip (<i> \psi R)} \right) \\
    <i + 1> \psi P &\equiv (<i + 1> \psi R) + 
      \left( \frac{(\transpose <i + 1> \psi R) \ip (<i + 1> \psi R)}{(\transpose <i> \psi R) \ip (<i> \psi R)} \right) \times
      (<i> \psi P) \\
    &\text{Result is $<i + 1> \psi X$.}
\end{align}
\subsubsection{Transpose}
Note that {\bf transpose} is an unnecessary operation on vectors in MoA (the typical issue of row- and column-vectors encountered in linear algebra is ignored. The reason why,
is based on the definition  of inner product, which  states that, 

{\em The shape of the result of the inner product, is the shape of the left argument, exclusive of  its last component on the right concatenated to the shape of the right argument, exclusive of its first component.  Also, the last component of the left argument, must be the same, 
as the first component of the right.}

\vspace {.1 in}
\noindent
So for example, with two vectors, the left argument $ {\vec w}$, and ${\vec v}$, the right argument, both have shape $<5>$, i.e., 
$\rho \; \vec w = <5>$ and $\rho \; \vec v = <5>$. Substituting
in what was said above: 
\begin{enumerate}
\item The last component of the  left argument, $ (-1) \take (\rho \; \vec w) = (-1) \take <5> = <5>$, and the first component of the  right arguement $ 1 \take (\rho \; \vec v) = 1 \take <5> = <5>$, are the same: 
\item Shape of the result: Everything but the last component of the left argument, which is the empty vector, $\Theta$, i. e. $ (-1) \drop (\rho \; \vec w) = (-1) \drop <5> = \Theta$, concatenated to everything but the first argument
of the right argument, also $\Theta$, i. e. $ 1 \drop (\rho \; \vec v) = 1 \drop <5> = \Theta$. 
Thus, $\Theta \cat \Theta$.  By definition of concatenation, the concatenation
of two empty vectors is $\Theta$, which is the shape of a scalar in MoA, that is,  what we know is the result of the  inner product of two vectors, or row vectors.

\end{enumerate}
\noindent
The {\bf inner product operator} behaves as expected on a pair of vector arguments, so each occurrence of transpose can be eliminated. Also note that, within a single iteration only two index values are used on the LHS of $\psi$: $<i + 1>$ and $<i>$. Without loss of generality, we can subtract $i$ from each index to reduce the size of working memory.
\begin{figure}[h]
\begin{align}
    <i - i + 1> \psi X &\equiv (<i - i> \psi X) + \left( \frac{(<i - i> \psi R) \ip (<i - i> \psi R)}{(<i - i> \psi P) \ip A \,\ip (<i - i> \psi P)} \right) \times (<i - i> \psi P) \\
    <i - i + 1> \psi R &\equiv (<i - i> \psi R) - \left( \frac{(<i - i> \psi R) \ip (<i - i> \psi R)}{(<i - i> \psi P) \ip A \,\ip (<i - i> \psi P)} \right) \times A \,\ip (<i - i> \psi P) \\
    <i - i + 1> \psi P &\equiv (<i - i + 1> \psi R) + 
      \left( \frac{(<i - i + 1> \psi R) \ip (<i - i + 1> \psi R)}{(<i - i> \psi R) \ip (<i - i> \psi R)} \right) \times
      (<i - i> \psi P) \\
    &\text{Result is $<i - i + 1> \psi X$.}
\end{align}
\caption{MoA: Subsequent Iterations - Transpose Removed}
\end{figure}
\subsubsection{Simplifying}
\begin{figure}[h]
Observing that $<i> - <i> \equiv <0>$ for all valid indices $<i>$, s.t. $ 0 \leq i <n$, we can simplify the expressions to
%\begin{figure}
\begin{align}
    <1> \psi X &\equiv (<0> \psi X) + \left( \frac{(<0> \psi R) \ip (<0> \psi R)}{(<0> \psi P) \ip A \,\ip (<0> \psi P)} \right) \times (<0> \psi P) \label{eq:simplify0}\\
    <1> \psi R &\equiv (<0> \psi R) - \left( \frac{(<0> \psi R) \ip (<0> \psi R)}{(<0> \psi P) \ip A \,\ip (<0> \psi P)} \right) \times A \,\ip (<0> \psi P) \label{eq:simplify2}\\
    <1> \psi P &\equiv (<1> \psi R) + 
      \left( \frac{(<1> \psi R) \ip (<1> \psi R)}{(<0> \psi R) \ip (<0> \psi R)} \right) \times
      (<0> \psi P) \label{eq:simplify1}\\
    &\text{Result is $<1> \psi X$.}
\end{align}
\caption{MoA: Subsequent Iterations - Simplifed Indexing}
\end{figure}
% Theorem 1 of the definition of inner product tells us that
% \begin{equation}
%   \xi_l (\,_{op_0} \bullet _{op_1}) \xi_r \equiv
%     (\,_{\,_{op_0} red} \Omega _{<\delta \xi_r>}) \xi_l' \; op_1 \; \xi_r'
% \end{equation}
% where
% \begin{gather}
%   \vec{x} \equiv (\rho \xi_l) \cat (1 \drop \rho \xi_r) \\
%   \vec{y} \equiv (1 \drop \rho \xi_r) \cat (\rho \xi_l) \\
%   m \equiv \,^-1 + (\delta \xi_l) + (\delta \xi_r) \\
%   \xi_l' \equiv
%     ((-(\,^-1 + \delta \xi_r)) \; \theta \; \iota m) \transpose (\vec{y} \rho \xi_l) \\
%    \xi_r' \equiv \vec{x} \rho \xi_r
% \end{gather}
\newpage
\subsection{Psi Reduction}
\subsubsection{MoA Definition of Inner Product: n-d}
\noindent
To paraphrase the definition of inner product, given that $\xi_l$ denotes the left argument, and $\xi_r$ the right, providing $1 \leq \delta \xi_l$ and $1 \leq \delta \xi_r$, i.e. the dimensionality of both arguments must be
greater or equal to 1.
\begin{equation}
  q \equiv (\,^- 1 \take \rho \xi_l)[0] \equiv (1 \take \rho \xi_r)[0]
\end{equation}
q is a scalar, so we must index the one element vector produced by
-1 take of the left shape and 1 take of the right shape.
Thus, the shape of the result is defined by
\begin{equation}
  \rho \xi_l \,\ip \xi_r \equiv
    (\,^- 1 \drop \rho \xi_l) \cat (1 \drop \rho \xi_r)
\end{equation}
and for $0 \leq i < (\,^-1 \drop \rho \xi_l)[0]$ and $0 \leq j < (\, 1 \drop \rho \xi_r)[0]$
\begin{equation}
  (<i \; j>) \psi \xi_l (\,_{op_0} \bullet _{op_1}) \xi_r \equiv
    \,_{\,_{op_0} red} (<i> \psi \xi_l) \, op_1 \,
    ((\iota q) \,_{\cat} \Omega _{<0 \; 1>} <j>) \psi \xi_r
\end{equation}

\subsubsection{Applying the Definition of Inner Product}
\noindent
Before we begin, we observe that many of the pieces of the derivation are the same.
Consequently, we pick pieces, reduce to normal form, then put the pieces together to
show an ONF for the CG.
Applying this definition, we can start to reduce the terms containing relevant inner product applications
\begin{equation}
  q \equiv
    (\,^- 1 \take \rho R)[0] \equiv
    (\,^- 1 \take <2 \; n>)[0] \equiv
    <n>[0] \equiv n
\end{equation}
\begin{align}
  (<0> \psi R) \ip (<0> \psi R) &\equiv
  (\Theta, \Theta) \psi (<0> \psi R) \ip (<0> \psi R) \\ &\equiv 
    \,_{\,_{+} red} (\Theta \psi <0> \psi R) \times
    ((\iota q) \,_{\cat} \Omega _{<0 \; 1>} \Theta) \psi (<0> \psi R) \\ 
& \nonumber \mbox{Here we say, take each scalar, from the left argument $\iota q$},\\
& \nonumber \mbox{and concatenate to each vector of the right argument, $\Theta$.}\\
&\nonumber \mbox{$\Theta$ indexing any array is that array.}\\
&\equiv
    \,_{\,_{+} red} (<0> \psi R) \times
    ((\iota n) \,_{\cat} \Omega _{<0 \; 1>} \Theta) \psi <0> \psi R \\
&\nonumber \mbox{The result of applying Omega is the following.}\\
 &\equiv
    \,_{\,_{+} red} (<0> \psi R) \times
      \begin{bmatrix}
        0 \\ \vdots \\ n - 1
      \end{bmatrix}
      \psi <0> \psi R \\
      \nonumber &\text{Note that the shape $<n \; 1>$, of the above array, selects every 
element of the } \\
      \nonumber &\text{right argument $<0> \psi R$, so we have} \\ &\equiv
    \,_{\,_{+} red} (<0> \psi R) \times <0> \psi R
\end{align}

From the definition of reduction, we know
\begin{equation}
  \,_+ red \; \vec{v} \equiv \sum_{i = 0}^{\,^-1 + \tau \vec{v}} \vec{v}[i]
\end{equation}

We can further simplify expression~(\ref{eq:simplify0}).
\begin{align}
  (<0> \psi R) \ip (<0> \psi R) &\equiv
    \sum_{j = 0}^{\,^-1 + \tau ((<0> \psi R) \times (<0> \psi R))} ((<0> \psi R) \times (<0> \psi R))[j] \\
    %&\equiv \sum_{j = 0}^{\,^-1 + \tau (<0> \psi R)} ((<0> \psi R) \times (<0> \psi R))[j] \\ &\equiv
    %\sum_{j = 0}^{n - 1} ((<0> \psi R) \times (<0> \psi R))[j] \\ &\equiv
    \nonumber \text{and since }
      & \,^-1 + \tau ((<0> \psi R) \times (<0> \psi R)) \equiv
      \,^-1 + \tau (<0> \psi R) \equiv
      n - 1 \\ &\equiv
    \sum_{j = 0}^{n - 1} (<0> \psi R)[j] \times (<0> \psi R)[j] \\ 
&\nonumber \mbox{In MoA, $\Ravel$ flattens any array, in various orderings, e.g. row major.}
\\& \nonumber \mbox{by PCT, which flattens contiguous row(s) of R based on ordering}\\
&\equiv
    \sum_{j = 0}^{n - 1} (\Ravel R)[(0 \times \pi (\,1 \take \rho R)) + j] \times (\Ravel R[(0 \times \pi (\,1 \take \rho R)) + j]) \\ &\equiv
    %\sum_{j = 0}^{n - 1} (\Ravel R)[j] \times (\Ravel R)[j] \\ &\equiv
    \sum_{j = 0}^{n - 1} (\Ravel R)[j]^2
\end{align}

\subsubsection{Another Common Expression}
The only difference in the following is $<1> \psi R$.
\noindent
Similarly, we can simplify expression ~(\ref{eq:simplify1}).
\begin{align}
  (<1> \psi R) \ip (<1> \psi R) &\equiv
    \sum_{j = 0}^{n - 1} (<1> \psi R)[j] \times (<1> \psi R)[j] \\ &\equiv
    \sum_{j = 0}^{n - 1} (\Ravel R)[(1 \times \pi (\,1 \take \rho R)) + j] \times (\Ravel R[(1 \times \pi (\,1 \take \rho R)) + j]) \\ &\equiv
    \sum_{j = 0}^{n - 1} (\Ravel R)[n + j]^2
\end{align}
\subsubsection{Final Common Section}
Now, we reduce the remaining inner product term in expressions ~(\ref{eq:simplify0}) and
~(\ref{eq:simplify2}). $\forall \;\;i, j\;\;\mbox{s.t.} \;\;0 \leq i < (\rho A)[0]\;\;\mbox{and} \;\;\;
0 \leq j < (\rho A)[1]$
\begin{align}
  (<0> \psi P) \ip A \,\ip (<0> \psi P) &\equiv
    (<0> \psi P) \ip (A \,\ip (<0> \psi P)) \\  &\equiv
    (<0> \psi P) \ip ((\Theta, \Theta) \psi A \,\ip (<0> \psi P)) \\  &\equiv
 % (<0> \psi P) \ip (\,_{\,_{+} red} (\Theta \psi A) \times <0> \psi P) \\ &\equiv
(<0> \psi P) \ip (\,_{\,_{+} red} (\Theta \psi A) _{\times} \Omega _{<1\;0>} <0> \psi P) \\ &\equiv
(<0> \psi P) \ip \sum_{j = 0}^{n - 1} (<j> \psi A) \times (<0> \psi P)[j] \\ &\equiv
    \nonumber (<0> \psi P) \ip \sum_{j = 0}^{n - 1} (\Ravel A)[(i \times \pi (\,1 \take \rho A)) + j] \times \\
      &\quad (\Ravel P)[(0 \times \pi (\,1 \take \rho P)) + j] \\ &\equiv
    (<0> \psi P) \ip \sum_{j = 0}^{n - 1} (\Ravel A)[(i \times n) + j] \times (\Ravel P)[j] \\ &\equiv
    \sum_{i = 0}^{n - 1} (\Ravel P)[i] \times \sum_{j = 0}^{n - 1} (\Ravel A)[(i \times n) + j] \times (\Ravel P)[j] \\ &\equiv
    \sum_{i = 0}^{n - 1} \sum_{j = 0}^{n - 1} (\Ravel P)[i] \times (\Ravel A)[j + i \times n] \times (\Ravel P)[j]
\end{align}

\subsubsection{Putting it All Together}

\begin{figure}[h]
Applying these reductions to the iterative algorithm step, 
we obtain the new expressions for all $0 \leq k < n$ and $\forall \;\;i, j \;\;\mbox{s.t.} \;\;0 \leq i < (\rho A)[0]\;\;\mbox{and} \;\;\;
0 \leq j < (\rho A)[1]$.
\begin{align}
    (\Ravel X)[n + k] &\equiv (\Ravel X)[k] + (\Ravel P)[k] \times \frac{\sum_{j = 0}^{n - 1} (\Ravel R)[j]^2}{\sum_{i = 0}^{n - 1} \sum_{j = 0}^{n - 1} (\Ravel P)[i] \times (\Ravel A)[j + i \times n] \times (\Ravel P)[j]} \\
    \nonumber (\Ravel R)[n + k] &\equiv (\Ravel R)[k] - \frac{\sum_{j = 0}^{n - 1} (\Ravel R)[j]^2}{ \sum_{i = 0}^{n - 1} \sum_{j = 0}^{n - 1} (\Ravel P)[i] \times (\Ravel A)[j + i \times n] \times (\Ravel P)[j]} \times \\ & \sum_{j = 0}^{n - 1} (\Ravel A)[(k \times n) + j] \times (\Ravel P)[j] \\
    (\Ravel P)[n + k] &\equiv (\Ravel R)[n + k] + 
      (\Ravel P)[k] \times \frac{\sum_{j = 0}^{n - 1} (\Ravel R)[n + j]^2 }{\sum_{j = 0}^{n - 1} (\Ravel R)[j]^2} \\
    &\text{Result is $(\Ravel X)[n + k]$.}
\end{align}
\caption{MoA: Subsequent Iterations - Psi Reduced, The ONF}
\end{figure}
\begin{figure}[h]
\subsection{Complete Iterative Solver, The MoA ONF}

%\begin{figure}[h]
For all $0 \leq k < n$, set the initial values
\begin{align}
  (\Ravel X)[k] &\equiv 0 \quad \text{(or better guess if available)} \\
  (\Ravel P)[k] &\equiv (\Ravel R)[k] \equiv (\Ravel \vec{b})[k] - \sum_{j = 0}^{n - 1} (\Ravel A)[(k \times n) + j] \times (\Ravel X)[j]
\end{align}

Then repeat the following steps:

Update the guess
\begin{multline}
    (\Ravel X)[n + k] \equiv (\Ravel X)[k] + (\Ravel P)[k] \times \\
      \left( \sum_{j = 0}^{n - 1} (\Ravel R)[j]^2 \right) /
      \left( \sum_{i = 0}^{n - 1} \sum_{j = 0}^{n - 1} (\Ravel P)[i] \times (\Ravel A)[j + i \times n] \times (\Ravel P)[j] \right)
\end{multline}

Calculate the new residual
\begin{multline}
    \nonumber (\Ravel R)[n + k] \equiv (\Ravel R)[k] -
      \left( \sum_{j = 0}^{n - 1} (\Ravel R)[j]^2 \right) /
      \left( \sum_{i = 0}^{n - 1} \sum_{j = 0}^{n - 1} (\Ravel P)[i] \times (\Ravel A)[j + i \times n] \times (\Ravel P)[j] \right) \times \\  \sum_{j = 0}^{n - 1} (\Ravel A)[j + k \times n] \times (\Ravel P)[j]
\end{multline}

Check if the new residual meets the convergence criterion or if we have completed all $n$ iterations of the direct solution; stop iterating if so. If not, calculate the new value of $P$ and loop back to update the guess again.
\begin{multline}
    (\Ravel P)[n + k] \equiv (\Ravel R)[n + k] + 
      (\Ravel P)[k] \times
      \left( \sum_{j = 0}^{n - 1} (\Ravel R)[n + j]^2 \right) / 
      \left( \sum_{j = 0}^{n - 1} (\Ravel R)[j]^2 \right)
\end{multline}

When complete, the solution is $(\Ravel X)[n + k]$. Note that this definition can be readily translated to a C-like language by replacing each raveled array with a pointer to an address in memory and using the square brackets as the C array dereference operator.
\caption{Complete Solver: MoA ONF} 
\end{figure}
\newpage
\subsubsection{Example}

We consider a linear system $A \ip \vec{x} \equiv \vec{b}$ given by
\begin{equation}
  A \ip \vec{x} \equiv
    \begin{bmatrix}
      4 & 1 \\ 1 & 3
    \end{bmatrix}
    \,\ip <x_0 \; x_1> \equiv
    <1 \; 2>
\end{equation}
This system differs from the linear algebra formulation by removing details about row and column vectors that are unnecessary for finding a solution.
\noindent
While we could start with any guess without loss of generality, we will consider
\begin{equation}
  \vec{x} \equiv <x_0 \; x_1> \equiv <2 \; 1>
\end{equation}
\noindent
The remaining initial values are then
\begin{align}
  (\Ravel P)[k] \equiv (\Ravel R)[k] &\equiv <1 \; 2>[k] - \sum_{j = 0}^{1} <4 \; 1 \; 1 \; 3>[j + k \times 2] \times <2 \; 1>[j] \\
  &\equiv <1 \; 2>[k] - (<4 \; 1 \; 1 \; 3>[0 + k \times 2] \times <2 \; 1>[0]) + (<4 \; 1 \; 1 \; 3>[1 + k \times 2] \times <2 \; 1>[1]) \\
  &\equiv <1 \; 2>[k] - (<4 \; 1 \; 1 \; 3>[k \times 2] \times 2) + (<4 \; 1 \; 1 \; 3>[1 + k \times 2])
\end{align}
\noindent
Since $0 \leq k < n$, we can expand easily for all values $k$
\begin{align}
  \nonumber (\Ravel P)[0] \equiv (\Ravel R)[0] &\equiv <1 \; 2>[0] - (<4 \; 1 \; 1 \; 3>[0 \times 2] \times 2) + (<4 \; 1 \; 1 \; 3>[1 + 0 \times 2]) \\
  &\equiv 1 - (4 \times 2) + 1 \equiv \,^-8 \\ \nonumber & \text{(Remember, by default, without any operation hierarchy,
MoA operations are evaluated from right to left)}\\
  \nonumber (\Ravel P)[1] \equiv (\Ravel R)[1] &\equiv <1 \; 2>[1] - (<4 \; 1 \; 1 \; 3>[1 \times 2] \times 2) + (<4 \; 1 \; 1 \; 3>[1 + 1 \times 2]) \\
  &\equiv 2 - (1 \times 2) + 3 \equiv \,^-3
\end{align}
So our initial condition is
\begin{equation}
0 \psi P \equiv 0 \psi R \equiv < \,^-8 \; \,^-3>
\end{equation}
\noindent
We now evaluate the next approximate solution by solving for $<1> \psi X$ for all valid $k$ and find
\begin{multline}
(\Ravel X)[1 + k] \equiv (\Ravel X)[k] + (\Ravel P)[k] \times \\
      \left( \sum_{j = 0}^{1} (\Ravel R)[j]^2 \right) /
      \left( \sum_{i = 0}^{1} \sum_{j = 0}^{1} (\Ravel P)[i] \times (\Ravel A)[j + i \times 2] \times (\Ravel P)[j] \right)
\end{multline}
\noindent
Substituting in the calculated initial condition, we have
\begin{multline}
(\Ravel X)[1 + k] \equiv <2 \; 1>[k] + < \,^-8 \; \,^-3>[k] \times \\
      \left( \sum_{j = 0}^{1} < \,^-8 \; \,^-3>[j]^2 \right) /
      \left( \sum_{i = 0}^{1} \sum_{j = 0}^{1} < \,^-8 \; \,^-3>[i] \times <4 \; 1 \; 1 \; 3>[j + i \times 2] \times < \,^-8 \; \,^-3>[j] \right)
\end{multline}
which evaluates to
\begin{equation}
(\Ravel X)[1 + k] \equiv <2 \; 1>[k] + < \,^-8 \; \,^-3>[k] \times \frac{73}{331} \approx <0.2356 \; 0.3384>
\end{equation}
\noindent
We then compute the next residual vector $<1> \psi R$
\begin{multline}
    (\Ravel R)[2 + k] \equiv < \,^-8 \; \,^-3>[k] -
      \left( \sum_{j = 0}^{1} < \,^-8 \; \,^-3>[j]^2 \right) / \\
      \left( \sum_{i = 0}^{1} \sum_{j = 0}^{n - 1} < \,^-8 \; \,^-3>[i] \times <4 \; 1 \; 1 \; 3>[j + i \times 2] \times < \,^-8 \; \,^-3>[j] \right) \times \\  \sum_{j = 0}^{1} <4 \; 1 \; 1 \; 3>[j + k \times 2] \times < \,^-8 \; \,^-3>[j]
\end{multline}
\begin{multline}
    (\Ravel R)[2 + k] \equiv < \,^-8 \; \,^-3>[k] -
      \frac{73}{331} \times
      \sum_{j = 0}^{1} <4 \; 1 \; 1 \; 3>[j + k \times 2] \times < \,^-8 \; \,^-3>[j]
\end{multline}
\begin{align}
    (\Ravel R)[2 + 0] &\equiv < \,^-8 \; \,^-3>[0] -
      \frac{73}{331} \times
      \sum_{j = 0}^{1} <4 \; 1 \; 1 \; 3>[j + 0 \times 2] \times < \,^-8 \; \,^-3>[j] \\
    (\Ravel R)[2] &\equiv \,^-8 - \frac{73}{331} \times
      \sum_{j = 0}^{1} <4 \; 1 \; 1 \; 3>[j] \times < \,^-8 \; \,^-3>[j] \\
    &\equiv \,^-8 - \frac{73}{331} \times
      (<4 \; 1 \; 1 \; 3>[0] \times < \,^-8 \; \,^-3>[0]) + (<4 \; 1 \; 1 \; 3>[1] \times < \,^-8 \; \,^-3>[1]) \\
    &\equiv \,^-8 - \frac{73}{331} \times
      (4 \times \,^-8) + (1 \times \,^-3) \\
    &\equiv \,^-8 - \frac{73}{331} \times \,^-35 \\
    &\equiv \,^-0.2810
\end{align}
\begin{align}
    (\Ravel R)[2 + 1] &\equiv < \,^-8 \; \,^-3>[1] -
      \frac{73}{331} \times
      \sum_{j = 0}^{1} <4 \; 1 \; 1 \; 3>[j + 1 \times 2] \times < \,^-8 \; \,^-3>[j] \\
    (\Ravel R)[3] &\equiv \,^-3 - \frac{73}{331} \times
      \sum_{j = 0}^{1} <4 \; 1 \; 1 \; 3>[j + 2] \times < \,^-8 \; \,^-3>[j] \\
    &\equiv \,^-3 - \frac{73}{331} \times
      (<4 \; 1 \; 1 \; 3>[2] \times < \,^-8 \; \,^-3>[0]) + (<4 \; 1 \; 1 \; 3>[3] \times < \,^-8 \; \,^-3>[1]) \\
    &\equiv \,^-3 - \frac{73}{331} \times
      (1 \times \,^-8) + (3 \times \,^-3) \\
    &\equiv \,^-3 - \frac{73}{331} \times \,^-17 \\
    &\equiv 0.7492
\end{align}
\noindent
The new residual vector is now $<1> \psi R \equiv <\,^-0.2810 \;  0.7492>$. Assuming this does not meet our convergence criterion, we will continue to calculate $<1> \psi P$.
\begin{align}
  \nonumber (\Ravel P)[2 + k] &\equiv <\,^-8 \; \,^-3 \; \,^-0.2810 \;  0.7492>[2 + k] + 
      < \,^-8 \; \,^-3>[k] \times \\
      &\left( \sum_{j = 0}^{1} <\,^-8 \; \,^-3 \; \,^-0.2810 \;  0.7492>[2 + j]^2 \right) / 
      \left( \sum_{j = 0}^{1} <\,^-8 \; \,^-3 \; \,^-0.2810 \;  0.7492>[j]^2 \right) \\
  \nonumber (\Ravel P)[2] &\equiv <\,^-8 \; \,^-3 \; \,^-0.2810 \;  0.7492>[2] + 
      < \,^-8 \; \,^-3>[0] \times \\
      &\left( \sum_{j = 0}^{1} <\,^-8 \; \,^-3 \; \,^-0.2810 \;  0.7492>[2 + j]^2 \right) / 
      \left( \sum_{j = 0}^{1} <\,^-8 \; \,^-3 \; \,^-0.2810 \;  0.7492>[j]^2 \right) \\
      &\equiv \,^-0.2810 + \,^-8 \times
      \left( \sum_{j = 0}^{1} <\,^-0.2810 \;  0.7492>[j]^2 \right) / 
      \left( \sum_{j = 0}^{1}<\,^-8 \; \,^-3>[j]^2 \right) \\
      &\equiv \,^-0.2810 + \,^-8 \times .6403 / 73 \\
      &\equiv \,^-0.3512 \\
  \nonumber (\Ravel P)[3] &\equiv <\,^-8 \; \,^-3 \; \,^-0.2810 \;  0.7492>[3] + 
      < \,^-8 \; \,^-3>[1] \times \\
      &\left( \sum_{j = 0}^{1} <\,^-8 \; \,^-3 \; \,^-0.2810 \;  0.7492>[2 + j]^2 \right) / 
      \left( \sum_{j = 0}^{1} <\,^-8 \; \,^-3 \; \,^-0.2810 \;  0.7492>[j]^2 \right) \\
      &\equiv 0.7492 + \,^-3 \times
      \left( \sum_{j = 0}^{1} <\,^-0.2810 \;  0.7492>[j]^2 \right) / 
      \left( \sum_{j = 0}^{1}<\,^-8 \; \,^-3>[j]^2 \right) \\
      &\equiv 0.7492 + \,^-3 \times .6403 / 73 \\
      &\equiv 0.7229
\end{align}
\noindent
So, $<1> \psi P \equiv <\,^-0.3512 \; 0.7229>$. Now we can iterate with the new values of $X$, $R$, and $P$ until a satisfactory solution is obtained.

\section{Translating to Pseudocode}
Note that we have switched back to using an assignment operator ``:='' to mimic execution behavior. Recall,   that this definition can be readily translated to a C-like language by replacing each raveled array with a pointer to an address in memory, denoted by {\bf X}, {\bf R}. and {\bf P},  and using the square brackets as the C array dereference operator.

\subsection{The CG: ONF to Pseudocode}
\begin{figure}[h]
For $0 \leq k < n$:
\begin{align}
  \mathbf{X}[k] &:= 0 \quad \text{(or better guess if available)} \\
  \mathbf{P}[k] &:= \mathbf{R}[k] :=  \vec{b}[k] - \sum_{j = 0}^{n - 1} \mathbf{A}[(k \times n) + j] \times \mathbf{X}[j]
\end{align}

While not converged:
\begin{equation}
    \mathbf{X}[n + k] := \mathbf{X}[k] + \mathbf{P}[k] \times
      \left( \sum_{j = 0}^{n - 1} \mathbf{R}[j]^2 \right) /
      \left( \sum_{i = 0}^{n - 1} \sum_{j = 0}^{n - 1} \mathbf{P}[i] \times \mathbf{A}[j + i \times n] \times \mathbf{P}[j] \right)
\end{equation}
\begin{equation}
    \nonumber \mathbf{R}[n + k] \equiv \mathbf{R}[k] -
      \left( \sum_{j = 0}^{n - 1} \mathbf{R}[j]^2 \right) /
      \left( \sum_{i = 0}^{n - 1} \sum_{j = 0}^{n - 1} \mathbf{P}[i] \times \mathbf{A}[j + i \times n] \times \mathbf{P}[j] \right) \times \sum_{j = 0}^{n - 1} \mathbf{A}[j + k \times n] \times \mathbf{P}[j]
\end{equation}

Exit now if converged.
\begin{equation}
    \mathbf{P}[n + k] \equiv \mathbf{R}[n + k] + 
      \mathbf{P}[k] \times
      \left( \sum_{j = 0}^{n - 1} \mathbf{R}[n + j]^2 \right) / 
      \left( \sum_{j = 0}^{n - 1} \mathbf{R}[j]^2 \right)
\end{equation}
\begin{align}
  \mathbf{X}[k] &:= \mathbf{X}[n + k] \\
  \mathbf{R}[k] &:= \mathbf{R}[n + k] \\
  \mathbf{P}[k] &:= \mathbf{P}[n + k]
\end{align}

Repeat.

Return $\mathbf{X}[n + k]$
\caption{The CG: ONF to Pseudocode}
\end{figure}
\subsubsection{Implementations from the ONF: Pseudocode}
We now have a design for hardware or software. 

%\input{notes-lowering}
%\input{code-frags}
%% The bibliography should be embedded for final submission.
%% \begin{thebibliography}{}
%% \softraggedright

%% \bibitem[Smith et~al.(2009)Smith, Jones]{smith02}
%% P. Q. Smith, and X. Y. Jones. ...reference text...

%% \end{thebibliography}
\newpage
\addcontentsline{toc}{section}{References}
\bibliography{paul_lenore}
\bibliographystyle{plain}

\end{document}